\documentclass[conference]{IEEEtran}

\usepackage[utf8]{inputenc}
\usepackage{amsmath, amssymb}
\usepackage{cite}
\usepackage{graphicx}
\graphicspath{{figures/}}
\usepackage{tablefootnote}
\usepackage{hyperref}

\title{Project Qualia: Recovering Experiential Music Structure from Session Co-occurrence Data}

\IEEEoverridecommandlockouts
\author{%
\IEEEauthorblockN{Nizam Mohammed\textsuperscript{*}}
\IEEEauthorblockA{Independent Researcher}
\and
\IEEEauthorblockN{Abu B. S. Rahman}
\IEEEauthorblockA{Department of Computer Science\\ Hampton University, VA 23669, USA}
\and
\IEEEauthorblockN{Dimuthu D. K. Arachchige}
\IEEEauthorblockA{Department of Computer Science\\ Hampton University, VA 23669, USA}
\thanks{\textsuperscript{*}Nizam Mohammed was with the School of Computing, DePaul University, Chicago, IL 60604, USA. He is now an independent researcher.}%
\thanks{Some of this work is intended for submission to an IEEE venue for a possible publication.}%
}

\begin{document}

\maketitle

\begin{abstract}
This report presents results from Project Qualia, an ongoing effort to determine whether experiential similarity between songs, a structure not captured by genre or metadata taxonomies, can be recovered from real listening behavior. We constructed a large-scale dataset of listening sessions, comprising 1.29 billion scrobbles collected from 9,396 users via the Last.fm API and reduced through a preprocessing pipeline to 531.6 million training scrobbles across 28.6 million sessions. On this corpus, we trained a skip-gram Word2Vec model (Song2Vec), treating each session as a sentence and each track as a token. As anticipated, the resulting embedding space was dominated by artist identity, a consequence of single-artist runs within sessions. To test for a subtler, artist-independent signal, we developed an artist-residual procedure: subtracting each artist's centroid from its tracks' embeddings and evaluating whether the remainder retained structure. Mean cross-artist cosine similarity fell from 0.2487 in raw embedding space to 0.0005 in residual space, yet 4,577 cross-artist track pairs retained cosine similarity $\ge 0.70$ in residual space, forming coherent genre- and era-based clusters, including trip-hop, 1990s grunge, 2020 mainstream pop, and cross-composer classical piano pairs at cosine similarity up to 0.95. These results confirm that the training data contains experiential structure independent of artist identity, establishing an empirical basis for an architecture designed to learn this experiential layer directly.
\end{abstract}


\section{Introduction}
Modern music recommendation systems reflect decades of substantive engineering progress: collaborative filtering~\cite{koren2009mf}, content-based audio analysis, deep learning over listening histories, and hybrid ranking systems refined against large-scale production datasets. These systems are sophisticated and, in most respects, accurate. Yet a consistent limitation persists across them: recommendations approximate what a listener wants to hear next without capturing it. Existing systems draw on genre, tempo, artist affinity, and co-listening statistics to produce outputs that are statistically plausible, but a gap remains between likely to be played next and what's consistent with the listener's current experience; a gap that has not closed appreciably across successive generations of these systems.

This project begins from the question of whether that gap reflects a limitation of available data or a limitation of system architecture. Yann LeCun's position paper, \textit{A Path Towards Autonomous Machine Intelligence}~\cite{lecun2022path}, motivates the latter interpretation. LeCun argues that intelligent systems require a predictive world model~\cite{ha2018worldmodels} operating over learned, abstract representations of their environment, rather than direct prediction over raw or shallow input features. Applied to music recommendation, this suggests that existing systems operate on proxies for listening experience, genre labels, artist metadata, engagement signals, rather than on a learned representation of the experience itself. If music possesses an underlying experiential structure distinct from, and irreducible to, artist identity or genre classification, a system trained to predict within that structure should in principle close the gap that current systems have not.

\begin{figure*}[t]
\centering
\includegraphics[width=\textwidth]{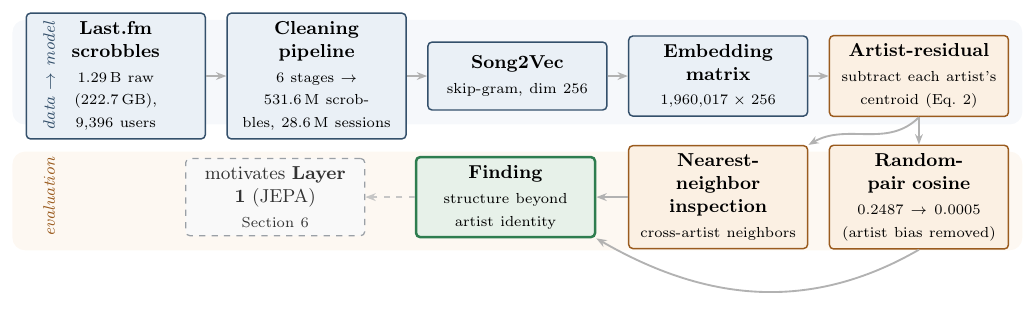}
\vspace{-12mm}
\caption{Overview of the study. Session-level Last.fm scrobbles are cleaned into a training corpus and used to train a skip-gram Song2Vec model, and the artist-residual transform of Section~3.3 is applied to the resulting embeddings. Two analyses of the residual space, a random-pair cosine comparison and a nearest-neighbor inspection, support the finding that the data carries structure beyond artist identity, which motivates the Layer~1 architecture in Section~6.}

\label{fig:pipeline}
\end{figure*}

Project Qualia treats this as a hypothesis subject to empirical test. The hypothesis is that a listening session, the sequence of songs a listener plays consecutively, constitutes a behavioral trace of an experiential structure not captured by any existing metadata schema, and that this structure is recoverable from data independent of any predictive architecture built to represent it. The project is accordingly structured in stages: first, establish whether this experiential signal exists in the data and can be recovered at all. Second, if it does, construct a predictive architecture, following LeCun's Joint Embedding Predictive Architecture (JEPA) framework, adapted to sequential listening data, capable of representing that structure directly. Each stage is defined to be falsifiable, and individual stages may not proceed as anticipated. The hypothesis under test in this report concerns the existence of the signal, not the eventual performance of any system built on it. This report addresses the first stage: constructing a dataset of sufficient scale to make the question answerable, and testing whether the hypothesized experiential signal is present in the resulting data. Figure~\ref{fig:pipeline} summarizes the study, from raw scrobbles to the residual-space finding.

\section{Data}
\subsection{Collection}
Existing public scrobble corpora, most notably Schedl's Last.fm-1B dataset \cite{schedl2016lfm}, are outdated and offer no visibility into the collection or preprocessing decisions behind them. For this reason, we collected a new dataset directly from the Last.fm public API.

Last.fm is a music-tracking service: connected clients record (``scrobble'') each track a user plays, along with a timestamp, artist name, track title, and, where available, a MusicBrainz ID (MBID),\footnote{MusicBrainz is a community-maintained open music encyclopedia and identifier service, \url{https://musicbrainz.org}.} a canonical identifier independent of how a given client reports the artist or track string. We targeted heavy users, defined as accounts with multi-year listening histories and tens to hundreds of thousands of scrobbles, on the premise that dense histories provide the richest session structure. Each user's complete scrobble history was retrieved through the Last.fm public API.

Collection yielded 1,292,098,037 scrobbles from 9,396 users, stored in a local SQLite database of approximately 222.7 GB. This database was compressed with zstd to 59 GB, verified with a SHA256 checksum, and archived; the uncompressed original was deleted following verification.

\subsection{Preliminary Sample Analysis}
Prior to operating on the full corpus, we conducted exploratory analysis on a random 200-user sample to identify structural issues that would need to be addressed at scale. Three findings from this sample determined the subsequent processing strategy.

First, string-based track identity is unreliable: the same recording appears under numerous surface forms (case variation, inconsistent ``ft.'' versus ``feat.'' conventions, inconsistently tagged live and alternate versions), inflating the apparent vocabulary by a factor of approximately 4.6 relative to MBID-based identity. MBIDs, however, are not present on every scrobble, so neither identity scheme is sufficient alone.

Second, the distribution of listening is heavily long-tailed: 77.9\% of unique tracks in the sample were played by exactly one user. A co-occurrence model can only learn a relationship between tracks it observes together, so this tail contributes no usable training signal regardless of how it is filtered downstream.

Third, session structure is recoverable using a fixed inter-track gap threshold. Applying a 30-minute gap rule produced a median intra-session gap of approximately 207 seconds, consistent with the length of an average track, and a mean session length of 23.9 tracks, with 53.2\% of sessions containing eight or more tracks. The sample also identified two contamination sources requiring explicit handling in later stages: 12 accounts (0.77\% of sample scrobbles) exhibiting mechanically uniform inter-track timing consistent with bot activity, and a K-Pop concentration; the BTS ecosystem alone accounted for 5.8\% of sample scrobbles, large enough to distort embedding density if left unaddressed.

\subsection{Analysis at Full Scale}
To operate on the full corpus at tractable speed, we converted the SQLite database to 188 ZSTD-compressed Parquet files (15 GB total, plus separate user-metadata and social-graph tables) and moved all subsequent analysis to DuckDB \cite{raasveldt2019duckdb}, which scans the full corpus in under a minute by reading only the required columns in parallel. All full-scale queries reported below completed in 169 seconds, and the corpus totals they produced are summarized in Table~\ref{tab:full_corpus_metrics}.

\begin{table}[htbp]
\centering
\caption{Full corpus summary metrics prior to pipeline filtering.}
\label{tab:full_corpus_metrics}
\begin{tabular}{lr}
\hline
\textbf{Metric} & \textbf{Value} \\
\hline
Total scrobbles (excluding timestamps predating 2002\tablefootnote{Last.fm was founded in 2002; timestamps predating the service's existence cannot reflect genuine scrobble activity and were excluded.}) & 1,252,761,419 \\
Unique string-based track identities & 30,443,573 \\
Unique MBIDs & 6,559,424 \\
Track count after MBID/string deduplication & 29,917,326 \\
Total sessions (30-minute gap rule) & 53,735,946 \\
\hline
\end{tabular}
\end{table}

Deduplication using MBIDs, with string keys as fallback, reduced the vocabulary by only 1.7\%, indicating that the majority of string-based inflation occurs among obscure, low-play-count tracks lacking an MBID rather than among high-frequency tracks. This motivated a listener-count threshold rather than string normalization as the primary vocabulary-reduction strategy: requiring a track be heard by at least ten distinct users reduces the vocabulary by 93\% (from 30.4 million to 2.2 million tracks) while discarding only 16.4\% of total scrobbles.

Two further patterns, identified at full scale, informed the cleaning pipeline described in Section 3. Repeat listening is substantial: 44.5\% of all scrobbles represent a user's fiftieth or later play of a given track, and 12.7\% are immediate same-track repeats within a session; both patterns encode individual listening habit rather than track-to-track relationship and would be expected to bias a co-occurrence model toward user-specific repetition rather than cross-track structure. Session length is also heavily skewed: sessions of 100 or more tracks constitute 3.1\% of all sessions but account for 38.6\% of all scrobbles. A subsequent gap-uniformity classification of these long sessions (detailed in Section 3) found 78\% to be natural human listening, 20.5\% ambiguous, and 1.5\% consistent with automated playback.

\section{Method}

\subsection{Cleaning Pipeline}
Guided by the characterization in Section 2, we applied a six-stage cleaning pipeline to the full corpus, executed once, in sequence, on the Parquet-converted data.

\paragraph{Stage 1: Session reconstruction.} Sessions were reconstructed using the 30-minute inter-track gap rule established during exploratory analysis (Section 2.2), with timestamps predating 2002 excluded per the threshold established in Section 2.3. This produced 53,735,946 raw sessions.

\paragraph{Stage 2: Automated-playback removal.} Sessions of 100 or more tracks were classified using a gap-uniformity heuristic: a session was flagged as automated playback if more than 80\% of its inter-track gaps fell within a $\pm 5$-second band of one another and no gap in the session exceeded five minutes, a timing signature consistent with continuous unattended playback and inconsistent with genuine human listening, which reliably contains pauses. Sessions meeting neither this criterion nor a stricter ``natural'' criterion (varied gap distribution, at least one pause exceeding five minutes) were classified as mixed and retained, since discarding them would have removed a large share of long sessions for uncertain benefit. This stage removed 12,697 sessions comprising 1,785,673 scrobbles. The heuristic is calibrated to detect uniform inter-track timing characteristic of automated or bot-driven scrobbling and does not identify algorithmically generated autoplay continuations, which reproduce naturalistic timing distributions; this limitation is addressed in Section 5.2.

\paragraph{Stage 3: Vocabulary thresholding.} We evaluated four candidate vocabulary thresholds, each defined by a minimum distinct-listener count and minimum total play count, against the tradeoff between vocabulary size and scrobble coverage (Table \ref{tab:vocab_thresholds}).

\begin{table*}[t]
\centering
\caption{Evaluation of candidate vocabulary reduction thresholds.}
\label{tab:vocab_thresholds}
\begin{tabular}{llrr}
\hline
\textbf{Threshold} & \textbf{Criterion} & \textbf{Vocabulary Size} & \textbf{Scrobble Coverage} \\
\hline
Conservative & $\ge 20$ users, $\ge 50$ plays & 1,092,916 & 77.9\% \\
Moderate & $\ge 10$ users, $\ge 25$ plays & 1,993,639 & 83.2\% \\
Aggressive & $\ge 5$ users, $\ge 10$ plays & 3,743,521 & 87.9\% \\
Ultra-aggressive & $\ge 3$ users, $\ge 5$ plays & 6,090,102 & 91.0\% \\
\hline
\end{tabular}
\end{table*}

We selected the moderate threshold, prioritizing a vocabulary small enough to receive adequate per-track co-occurrence signal during training over marginal coverage gains from looser thresholds. Applied to the corpus, this threshold retained 1,960,021 of 28,820,633 total unique tracks (6.80\% of raw vocabulary) and covered 84.1\% of scrobbles; the remaining 15.9\% (194,621,753 scrobbles) were mapped to an out-of-vocabulary token and excluded from training.

\paragraph{Stage 4: Consecutive-repeat collapse.} Within each session, immediate repeated plays of the same track (e.g., a track looped by the user) were collapsed to a single occurrence, removing a signal that reflects looping behavior rather than track-to-track co-occurrence. Scrobble count fell from 1,225,936,047 to 979,448,409, a reduction of 246,487,638 scrobbles (20.1\%).
\paragraph{Stage 5: Per-user, per-track play capping.} Each user's total plays of a given track were capped at 30, with the retained occurrences distributed evenly across that user's listening history, to prevent individual superfan behavior from dominating the co-occurrence statistics of a small number of tracks. This stage reduced the scrobble count from 979,448,409 to 591,679,449, a reduction of 387,768,960 scrobbles (39.6\%).

\paragraph{Stage 6: Final session filtering and sharding.} Sessions retaining fewer than five tracks after the preceding stages were discarded as insufficient for co-occurrence learning. The remaining sessions were partitioned into 189 Parquet shards. The resulting training corpus comprised 28,593,422 sessions and 531,629,984 scrobbles (mean session length 18.6 tracks, median 11), spanning a vocabulary of 1,960,021 tracks, at 3,747 MB on disk. 24.9\% of sessions met the single-artist-dominance criterion ($\ge 80\%$ of tracks by one artist) and were retained as valid single-artist listening rather than removed as redundant.

\subsection{Embedding Model}
We trained a skip-gram Word2Vec model~\cite{mikolov2013efficient} with negative sampling~\cite{mikolov2013distributed} via Gensim~\cite{rehurek2010gensim} on the 189 session shards, treating each session as a sentence and each track as a token: the same formulation used to learn word embeddings from text, applied here to sequences of listening behavior~\cite{barkan2016item2vec}.

For a center track $w_c$ and a context track $w_o$ occurring within the window, with $k$ negative samples drawn from a noise distribution $P_n$, training maximizes
\begin{equation}
\log \sigma(v'_{w_o} \cdot v_{w_c}) + \sum_{i=1}^{k} \mathbb{E}_{w_i \sim P_n}\left[\log \sigma(-\,v'_{w_i} \cdot v_{w_c})\right],
\end{equation}
where $v$ and $v'$ denote the input and output embeddings, respectively.

\begin{table}[htbp]
\centering
\caption{Skip-gram Word2Vec training hyperparameters.}
\label{tab:word2vec_hyperparams}
\begin{tabular}{lr}
\hline
\textbf{Hyperparameter} & \textbf{Value} \\
\hline
Embedding dimensionality & 256 \\
Context window & 10 \\
Minimum token count & 1 \\
Negative samples per positive example & 15 \\
Training epochs & 5 \\
Parallel workers & 8 \\
\hline
\end{tabular}
\end{table}

Trained with the hyperparameters in Table~\ref{tab:word2vec_hyperparams}, training produced a $1,960,017 \times 256$ float32 embedding matrix ($\sim 1.9$ GB), the underlying Gensim model file, and a track-to-index mapping.

\subsection{Artist-Residual Analysis (Experiment A)}
The central open question was whether proximity in the trained embedding space reflects genuine sonic or experiential similarity, or is dominated by artist identity, a plausible confound, since most sessions contain long single-artist runs (Section 3.1, Stage 6). We designed an artist-residual procedure to isolate these two components.

\begin{figure*}[t]
\centering
\includegraphics[width=\textwidth]{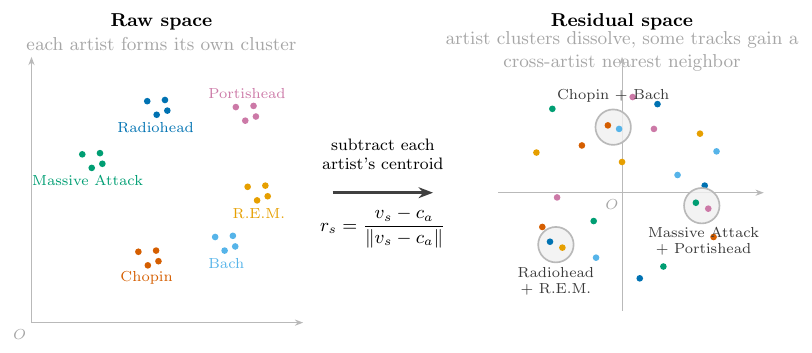}
\caption{The artist-residual procedure (Eq.~2). Points are colored by artist and carry no genre or audio information, and artist labels are used only to remove each artist's centroid. In raw space each artist forms its own cluster. After subtracting each artist's centroid and normalizing, most tracks remain diffuse while a minority of tracks by different artists become nearest cross-artist neighbors, shown here for three illustrative pairs with residual cosine 0.61 to 0.94. The panels are a schematic of the operation and are not a projection of the 256-dimensional data.}
\label{fig:concept}
\end{figure*}

For every artist with at least five tracks present in the vocabulary (52,445 artists, covering 1,778,648 tracks, or 90.7\% of the vocabulary), we computed an artist centroid as the mean of that artist's track embeddings. For each track by that artist, we subtracted the artist centroid from the track's embedding and $L_2$-normalized the result, producing a residual vector representing how that track diverges from its artist's typical embedding position. For an artist $a$ with track set $T_a$, the centroid $c_a$ and the residual $r_s$ for a track $s \in T_a$ with embedding $v_s$ are
\begin{equation}
c_a = \frac{1}{|T_a|} \sum_{t \in T_a} v_t, \qquad r_s = \frac{v_s - c_a}{\lVert v_s - c_a \rVert_2}.
\end{equation}
If the embedding space encodes only artist identity, residual vectors should be effectively random with respect to other artists' residuals; if it encodes an experiential structure independent of artist identity, residual vectors from different artists should show non-random alignment where the underlying tracks are genuinely similar in feel. Figure~\ref{fig:concept} illustrates the operation.

We evaluated this by comparing cosine similarity distributions between raw embeddings and residual embeddings across random cross-artist track pairs, and by retrieving nearest neighbors in residual space for a fixed set of well-known tracks to inspect cluster composition directly.

\section{Results}

\subsection{Artist Identity Dominates Raw Embedding Space}
Across randomly sampled cross-artist track pairs, raw embeddings showed a mean cosine similarity of 0.2487. Since these pairs were drawn without regard to genre, era, or any other similarity criterion, a positive baseline of this magnitude indicates a systematic bias: the embedding space contains a shared directional component correlated with artist identity, present across the vocabulary rather than confined to any subset of tracks. This confirms the concern motivating Experiment A: co-occurrence learned primarily from single-artist session runs (Section 3.1) manifests as an artist-identity signal strong enough to dominate raw proximity.

\subsection{Residual Space Isolates a Second, Independent Signal}
After subtracting each artist's centroid and $L_2$-normalizing (Section 3.3), the same cross-artist comparison, over 100,000 randomly sampled pairs of tracks with above-median residual norm, yielded a mean cosine similarity of 0.0005 (SD 0.086), a reduction consistent with near-complete removal of the artist-identity bias identified in Section 4.1. Residual space is, on average, uncorrelated across artists. Sampling in this section is restricted to tracks whose residual norm exceeds the median, which excludes near-zero residuals whose orientation is unstable.

Against this near-zero baseline, 4,577 cross-artist track pairs among the ten nearest residual neighbors of 50,000 sampled tracks retained cosine similarity $\ge 0.70$ in residual space, a value roughly eight standard deviations above the baseline mean. Because the baseline is effectively random, these pairs represent structure that cannot be attributed to artist identity: two tracks by different artists, positioned close together after each track's own artist-typical direction has been removed.

Residual norms across the 1,778,648 qualifying tracks (mean 2.6457, SD 0.7022, median 2.6447, P5--P95 range 1.51--3.81) varied by how atypical a track was relative to its artist's centroid. Tracks with the highest residual norms were, on inspection, soundtrack contributions or genre-atypical collaborations; for example Kodak Black's ``Angel Pt. 1'' (norm 6.06) and Paul McCartney's ``Love Song to the Earth'' (5.86). Tracks with the lowest residual norms belonged to artists with narrow, sonically uniform catalogs, such as ambient/sleep-sound acts, where individual tracks are near-indistinguishable from the artist's centroid.

\subsection{Cluster Composition in Residual Space}
\begin{figure*}[t]
\centering
\includegraphics[width=\textwidth]{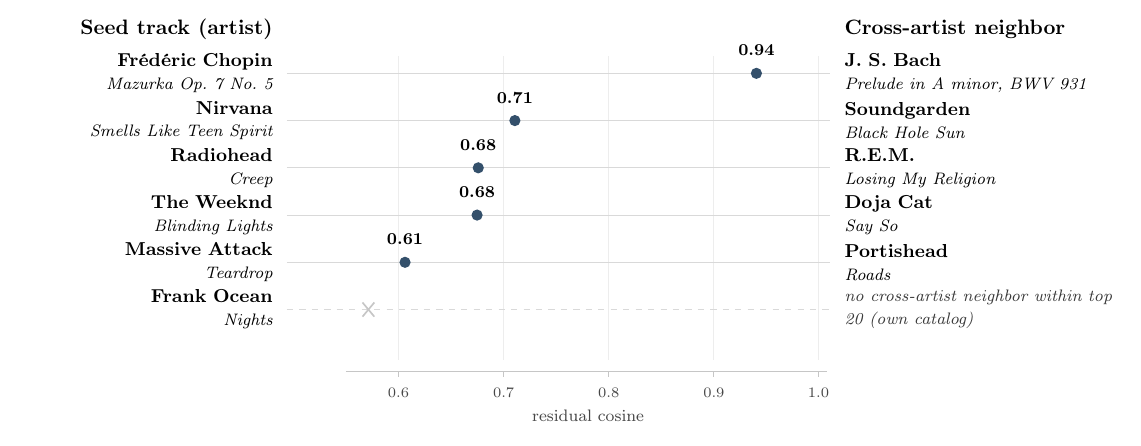}
\caption{Residual-space nearest cross-artist neighbors for representative seed tracks, by cosine similarity. In raw space each pop/rock seed's nearest neighbors are its own artist. Each cross-artist neighbor shown here appears at rank 5 to 8 among all residual neighbors, and at rank 2 for Chopin, below a same-artist core. Frank Ocean is included as a case where no cross-artist neighbor appears within the top 20. Examples are hand-selected from the 20 seed tracks inspected in Section~4.3 and are not a systematic sample.}
\label{fig:ledger}
\end{figure*}

To evaluate whether the cross-artist residual signal corresponds to interpretable structure, we retrieved the top 15 residual-space nearest neighbors for 20 well-known seed tracks. Two patterns emerged. Figure~\ref{fig:ledger} shows representative examples.

\paragraph{Cross-artist clustering by genre and era.} For seed tracks drawn from genres with clear stylistic contemporaries, residual-space neighbors beyond the seed artist's own catalog consistently resolved to genre- and era-coherent clusters:

\begin{itemize}
    \item \textbf{Radiohead, ``Creep'':} Cross-artist neighbors from rank 5 included R.E.M.'s ``Losing My Religion'' (0.676), Beck's ``Loser'' (0.645), Red Hot Chili Peppers' ``Under the Bridge'' (0.619), Counting Crows' ``Mr. Jones'' (0.616), and The Police's ``Every Breath You Take'' (0.615); a 1990s alternative-rock cluster.
    \item \textbf{Massive Attack, ``Teardrop'':} Cross-artist neighbors included Portishead's ``Roads'' (0.606) and ``Glory Box'' (0.602), Moby's ``Porcelain'' (0.598) and ``Natural Blues'' (0.555), and Morcheeba's ``The Sea'' (0.537); a trip-hop cluster centered on the genre's two defining acts.
    \item \textbf{Nirvana, ``Smells Like Teen Spirit'':} Cross-artist neighbors included Soundgarden's ``Black Hole Sun'' (0.711) and Pearl Jam's ``Jeremy'' (0.631); a grunge cluster.
    \item \textbf{The Weeknd, ``Blinding Lights'':} Beyond four same-artist neighbors, the remainder of the top 15 was almost entirely cross-artist: Doja Cat, Post Malone, Future, Harry Styles, Justin Bieber, Dua Lipa; resolving to a 2020 mainstream-pop cluster.
    \item \textbf{Fr{\'e}d{\'e}ric Chopin, Mazurkas, Op. 7 No. 5:} The closest cross-artist residual-space neighbor, at cosine 0.941, was Johann Sebastian Bach's Prelude in A minor, BWV 931 (a same-artist Chopin prelude ranks just above it at 0.955), two centuries apart in composition, adjacent in residual space. Comparable cross-composer classical piano pairs (Chopin--Bach at 0.937, 0.933, 0.924; Bizet--Brahms at 0.932; Saint-Sa{\"e}ns--Grieg at 0.946 and 0.935) suggest the residual dimension here tracks tonal character and pianistic texture rather than composer identity.
\end{itemize}

\paragraph{Artist-internal clustering where catalog identity is unusually cohesive.} A minority of seed tracks did not resolve to cross-artist clusters. Frank Ocean's ``Nights'' retained all 15 neighbors within his own catalog, at cosine similarities of 0.775--0.920, consistent with a production style unified enough that residual subtraction does not separate his tracks from one another. Taylor Swift's ``All Too Well'' and Kendrick Lamar's ``HUMBLE.'' showed a similar, if less extreme, pattern (all 15, and 13 of 15, same-artist neighbors respectively). Billie Eilish's ``when the party's over'' showed the inverse: only 2 of 15 neighbors were her own tracks, with the remainder (Halsey, Khalid, Kodak Black) placing her within a broader late-2010s alternative-pop cluster rather than a distinctive artist identity.

\subsection{Interpretation}
Experiment A was designed as a validation step, not the recommendation model itself. Before committing to a more complex predictive architecture, it is necessary to establish that the training data contains the signal such an architecture would need to learn. Word2Vec is well suited to this check because it is inexpensive to train, well understood, and directly interpretable: if the corpus contained no experiential structure independent of artist identity, a more sophisticated model would not manufacture that structure from the same data. The experiment was therefore constructed to be falsifiable: a near-zero count of high-similarity cross-artist pairs in residual space would have indicated that the 531.6 million-scrobble training corpus encodes artist affinity and little else, and the project's premise would not have survived contact with the data.

That did not happen. The result reported in Sections 4.2--4.3 (4,577 cross-artist pairs at cosine similarity $\ge 0.70$ against a random baseline of 0.0005, resolving into genre- and era-coherent clusters including trip-hop, grunge, 90s alternative rock, 2020 mainstream pop, and cross-composer classical piano without any genre label supplied during training) establishes two things with the confidence the experiment was designed to produce.

First, the dataset carries the intended signal. The listening behavior of 9,396 heavy users, collapsed to session-level co-occurrence, contains enough structure to recover cross-artist sonic and stylistic relationships that a purely artist-identity-driven signal could not produce.

Second, Word2Vec identifies the ceiling of what a pure co-occurrence model can do here. Skip-gram Word2Vec has no notion of sequence order beyond a symmetric context window, separates artist identity from experiential similarity only by post hoc residual subtraction, and has no representation of a session's structure as a whole. The artist-residual procedure worked as a diagnostic precisely because it was applied after training, to embeddings that had no access to that distinction during learning. This is enough to confirm that a signal exists; it is not enough to build a model intended to represent that signal directly, and Word2Vec was never proposed as that model.

The result reported here confirms readiness for the next stage: it does not complete the project. The data justifies investment in a predictive architecture of the kind described in Section 6, designed to represent artist identity and experiential similarity as separable components from the outset, rather than recovering their separation after training.

\section{Discussion \& Limitations}

\subsection{What This Result Does and Doesn't Establish}
The result in Section 4 addresses a narrow but necessary question: does session-level listening co-occurrence, at the scale collected here, encode anything about music beyond artist identity? It does. That finding is a precondition for the architecture proposed in Section 6, not evidence for that architecture's eventual performance: the residual analysis validates the presence of a signal, not the correctness of any particular way of modeling it.

It is also worth being precise about what ``experiential similarity'' means here, since the term is doing significant work. The clusters recovered in residual space (trip-hop, grunge, 90s alternative rock, 2020 mainstream pop, cross-composer classical piano) are genre- and era-coherent, consistent with the hypothesis that the model is recovering something about how tracks sound or feel. But genre coherence and experiential similarity are distinct claims: a cluster could form because tracks share production era, instrumentation, or tempo, properties that overlap with genre by construction without being identical to it. The residual analysis demonstrates the signal is not reducible to artist identity; distinguishing whether it reflects a listener's sense of shared ``feel,'' specifically, versus a narrower or different kind of acoustic or contextual similarity, was outside the scope of the current analysis and remains open.

The evidence itself rests on two forms of validation: an aggregate, threshold-based statistic (4,577 cross-artist pairs at cosine $\ge 0.70$ against a 0.0005 random baseline) and a qualitative nearest-neighbor inspection of 20 seed tracks. The seed-track inspection was illustrative rather than drawn through a blind or systematic sampling procedure, which limits how strongly the specific examples in Section 4.3 can be read as representative of the full 4,577-pair set rather than as favorable instances within it. No held-out quantitative evaluation, such as a human-judged similarity benchmark or a downstream task with a measurable success criterion, was used to validate cluster quality. That is consistent with this stage's role as an existence check rather than a model evaluation, but it means the results support the claim that a signal exists, not a claim about how strong or reliable that signal is for a downstream recommendation task.

The artist-residual method itself carries an assumption: that an artist's typical embedding position is well approximated by the linear mean of their tracks' embeddings, and that subtracting this mean isolates identity from feel. This assumption held well enough to produce the near-zero random baseline in Section 4.2, which is evidence in its favor, but it is a simple linear operation. Artists with unusually cohesive catalogs (Frank Ocean, Taylor Swift, Kendrick Lamar in the seed-track results) did not separate cleanly under this method. That may reflect a genuine property of those artists' output, or a limitation of centroid subtraction as a separation technique: the current analysis does not distinguish between the two.

\subsection{Autoplay and the Provenance of Session Structure}
The cleaning pipeline described in Section 3.1 does not differentiate scrobbles produced by client-side autoplay, such as a streaming service's automatic queue continuation or Last.fm's own radio function, from tracks selected directly by the listener. Autoplay-generated scrobbles exhibit naturalistic inter-track timing and are not distinguishable from manually selected tracks at the level of the API response. The gap-uniformity heuristic applied in Stage 2 is calibrated to detect mechanically uniform timing associated with bot activity and does not detect autoplay sequences of this kind. Consequently, the co-occurrence structure on which Song2Vec was trained may partially reflect the recommendation logic of the originating platforms rather than independent listener selection, a circularity described for music recommender systems generally by Schedl et al.~\cite{schedl2018challenges}.

The magnitude of this effect cannot be determined from the present dataset. The Last.fm API method used for collection, \texttt{user.getRecentTracks} (Section 2.1), does not return a field indicating playback source or context; accordingly, the proportion of autoplay-originated scrobbles cannot be estimated from data already collected. A direct estimate would require either a dataset collected with client-side instrumentation recording playback trigger, or an indirect method comparing session orderings against documented platform playlist and radio sequences. Neither approach was incorporated into the present collection design.

Certain findings in Section 4.3 provide limited evidence against the signal being attributable solely to autoplay. The correspondence between Chopin and Bach, for example, spans catalogs and eras unlikely to co-occur within a single platform's autoplay sequence. Conversely, the 2020 mainstream-pop cluster is the case in which autoplay-driven contamination is least distinguishable from genuine cross-artist similarity, given that commercial genre radio is itself constructed to group stylistically similar recordings. The present analysis does not adjudicate between these two explanations. Prior to training the Layer 1 architecture described in Section 6 at scale, restricting the residual analysis to sessions originating from clients without autoplay functionality is recommended as a direct test of this confound.

\subsection{Population and Sampling Limitations}
The dataset reflects Last.fm's heavy-user population specifically: accounts with multi-year histories and high scrobble counts, selected deliberately for session density (Section 2.1). This population is not representative of listeners generally. Heavy scrobblers skew toward users who actively curate and track their listening, which likely differs systematically from casual listening behavior in genre breadth, repeat-listening habits, and session structure. Whatever structure Song2Vec learned was learned from this specific population, and how well it generalizes beyond it is untested.

The dataset also carries a known concentration bias: the BTS ecosystem alone accounts for 5.7\% of the full corpus, and the broader K-Pop category (49 acts) for 6.5\%, driven by a comparatively small number of heavy listeners (356 users, 3.8\% of the sample, with BTS in their top 5 artists). This is large enough to influence embedding density in that region of the space, and similar, less visible concentration effects plausibly exist for other genres or fan communities without having been specifically checked for. The per-user, per-track play cap (Section 3.1, Stage 5) mitigates the effect of individual superfans on any single track, but does not address concentration arising from many distinct heavy listeners of the same broader artist ecosystem.

\subsection{Architectural Limitations of the Validation Model}
As established in Section 4.4, skip-gram Word2Vec was selected specifically because it is inexpensive, well understood, and interpretable: properties suited to a validation step, not properties that justify it as a terminal model. Its limitations for the stated goal are structural rather than incidental: it treats a session as an unordered bag of co-occurring tracks within a fixed symmetric window, with no representation of sequence order, no explicit mechanism for separating identity from feel during training, and no model of session-level structure beyond pairwise co-occurrence. The residual procedure in Section 3.3 works around the second limitation after the fact; it does not address the first or third. These are the specific gaps the architecture proposed in Section 6 is intended to close.

\section{Future Work}

\subsection{Overview}
The analysis in Sections 2 through 5 established that session-level listening co-occurrence, at the scale collected for this project, contains structure independent of artist identity. This section specifies the architecture that result motivates for the next phase. Everything in this section describes planned work: no component below has been implemented, and no results are reported for it.

\subsection{Proposed System Architecture}
We propose a five-layer architecture, in which each layer depends on the validated output of the layer preceding it (Table~\ref{tab:system_architecture}).

Development proceeds under a fixed constraint: a given layer is not designed until the layer beneath it has satisfied a predefined evaluation criterion. Under this constraint, Layers 2 through 5 are not yet designed and are stated here only as intended scope. Sections 6.3 through 6.5 concern Layer 1, the next phase of work.

\begin{table*}[t]
\centering
\caption{Proposed five-layer system architecture for Project Qualia.}
\label{tab:system_architecture}
\begin{tabular}{lp{14cm}}
\hline
\textbf{Layer} & \textbf{Function} \\
\hline
1: Experiential song embeddings & Learns a latent space in which geometric proximity encodes experiential interchangeability between songs. \\
2: Transition model & Given a listener's current state (a point in the Layer 1 space) and a song played, predicts the listener's resulting state. \\
3: Policy / RL agent~\cite{sutton2018reinforcement} & Uses the Layer 2 transition model to plan session-level sequences of songs rather than individual next-song selections. \\
4: User-specific adaptation & Adapts the pretrained Layers 1--3 to an individual listener via interaction data, following the two-stage pretraining-and-adaptation pattern reported for V-JEPA 2~\cite{assran2025vjepa}. \\
5: Reasoning interface & Translates natural-language listener intent into session-level decisions. \\
\hline
\end{tabular}
\end{table*}

\subsection{Proposed Layer 1 Objective and Theoretical Basis}
The Song2Vec model reported in Sections 3.2 and 3.3 is not the proposed architecture for Layer 1. Its skip-gram objective predicts a song's identity from surrounding context, and Section 4 showed the resulting embedding space to be dominated by artist identity, requiring a post hoc residual-subtraction procedure to isolate an independent signal. The Layer 1 objective under consideration instead follows the Joint Embedding Predictive Architecture (JEPA) framework proposed by LeCun~\cite{lecun2022path}, in which a model is trained to predict the representation of a masked portion of its input from surrounding unmasked context, with both prediction and target expressed in a shared latent space rather than as a reconstructed or classified observation.

This framework has prior implementations directly relevant to the proposed design. I-JEPA~\cite{assran2023self} applies this objective to static images, predicting the representations of masked image regions from visible context using a Vision Transformer encoder~\cite{dosovitskiy2021image}. V-JEPA and V-JEPA 2~\cite{bardes2024vjepa, assran2025vjepa} extend the same principle to video, additionally establishing the two-stage pretraining-then-adaptation procedure referenced for Layer 4 above. The Layer 1 design adapts this framework to sequential listening data: a session is represented as an ordered sequence of song tokens, and the model is trained to predict the representation of a masked contiguous block of songs from the remaining, unmasked positions in the session.

This requires three components with no analogue in the Song2Vec formulation: a context encoder, operating on the unmasked positions of a session and producing predicted representations at the masked positions; a target encoder, following the exponential-moving-average construction introduced by BYOL~\cite{grill2020byol} and used in I-JEPA and V-JEPA~\cite{assran2023self, bardes2024vjepa}, providing the representations the context encoder is trained to predict; and a predictor module, mapping context-encoder output to the target encoder's representation space.

Writing $E_\theta$ for the context encoder, $E_\xi$ for the exponential-moving-average target encoder, $P$ for the predictor, $M$ for the masked block, and $x_{\setminus M}$ for the unmasked context, the objective is
\begin{equation}
\mathcal{L} = \sum_{m \in M} \left\lVert P\left(E_\theta(x_{\setminus M})\right)_m - \operatorname{sg}\left(E_\xi(x)\right)_m \right\rVert_2^2,
\end{equation}
where $\operatorname{sg}$ is the stop-gradient operator and $\xi$ is updated as an exponential moving average of $\theta$.

Masking is applied to a contiguous block drawn from the middle of a session, rather than to the symmetric sliding window used by the Song2Vec skip-gram objective, requiring the model to predict an extended portion of a listening session rather than individual adjacent tokens.

The basis for adopting this objective follows from the result in Section 4.1: an objective defined over song identity is satisfiable by learning artist affinity alone, which required correction after training in Experiment A. An objective defined entirely in representation space, with no song-identity target, aims to avoid that confound during training rather than correcting for it afterward. Whether it does so in practice is the subject of the evaluation described in Section 6.5, not an established result.

\subsection{Unresolved Design Parameters}
Several components of the proposed Layer 1 model remain unresolved:
\begin{itemize}
    \item \textbf{Representation collapse.} A recognized failure mode of joint-embedding predictive objectives is representation collapse, in which an encoder minimizes the prediction loss by mapping all inputs to a constant or near-constant representation, satisfying the objective without learning meaningful structure. Two candidate mitigations are under consideration: variance-invariance-covariance regularization, as proposed by Bardes, Ponce, and LeCun under the name VICReg~\cite{bardes2021vicreg}, and contrastive training objectives~\cite{chen2020simclr}. No selection has been made.
    \item \textbf{Listener-specific variation.} Both the completed Song2Vec model and the proposed Layer 1 objective operate on session data pooled across users. Whether listener-specific context (mood, situational intent, personal listening history) requires an explicit latent variable at Layer 1, or can be treated as variation averaged out at corpus scale, is unresolved.
    \item \textbf{Acoustic feature integration.} The current Layer 1 formulation uses canonical track-identifier embeddings only, with no audio-derived features. Incorporating MERT~\cite{li2023mert}, a pretrained self-supervised model for music-audio representation, is under consideration as either an auxiliary input channel, a source of soft target labels, or a separate parallel pathway. No decision has been made among these options.
    \item \textbf{Encoder configuration.} Sequence length, attention mechanism, and positional encoding for the context and target encoders are not yet finalized. An initial working range under consideration is 4 to 6 transformer layers~\cite{vaswani2017attention}, 256 to 512 hidden dimensions, and attention windows of 20 to 50 songs.
    \item \textbf{Treatment of repeated listens.} The per-user, per-track play cap applied during Song2Vec preprocessing (Section 3.1, Stage 5) removed 39.6\% of scrobbles by limiting each user's plays of a given track to 30. Under a JEPA-style objective, individual repeat-listen events may constitute distinct behavioral contexts rather than redundant signal, which would make this cap unsuitable in its current form. One direction under consideration is to remove or substantially raise the cap and instead manage listener-level concentration through frequency-based subsampling at training time; this has not been finalized.
    \item \textbf{Album-block compression.} Whether to replace consecutive same-artist runs of four or more tracks with a single compressed token prior to training is proposed but not yet evaluated in either configuration.
\end{itemize}

\subsection{Proposed Evaluation Protocol}
Consistent with the layer-gating principle in Section 6.2, evaluation criteria for Layer 1 are defined in advance of training, along three axes:
\begin{enumerate}
    \item \textbf{Nearest-neighbor inspection.} The qualitative method of Experiment A (Section~3.3), applied to 20 seed tracks in Section~4.3, is extended to a fixed set of 50 seed tracks spanning multiple genres and moods.
    \item \textbf{Genre entropy within clusters.} Clusters exhibiting high genre entropy (containing tracks from multiple genres rather than one) are interpreted as evidence the model has organized tracks along a dimension other than genre.
    \item \textbf{Session-completion accuracy.} Held-out contiguous segments of real sessions are masked, and the model's ability to recover them is evaluated against three baselines: random selection, popularity-weighted selection, and genre-matched selection. Outperforming the genre-matched baseline is the deciding criterion, since a model that has learned genre as a proxy for experiential structure (without capturing structure beyond it) would be expected to match rather than exceed this baseline.
\end{enumerate}

Design of Layer 2 proceeds only if all three criteria are satisfied. If they are not, the next step is revision of the Layer 1 objective, preprocessing, or architecture, prior to any further work on Layers 2 and above.

\subsection{Scope of Present Claims}
The result reported in Section 4 justifies proceeding to the architecture and evaluation plan in this section; it does not establish that architecture's performance. Several questions remain open as of this report: whether a JEPA-style objective recovers experiential structure more effectively than the residual-corrected Song2Vec embeddings already do; whether Layer 1 representations, once trained, are suitable state representations for a Layer 2 transition model and Layer 3 policy (a use motivated by the JEPA and world-model literature cited above~\cite{lecun2022path, ha2018worldmodels, assran2025vjepa, assran2023self} but not evaluated on this dataset); and whether the cross-genre clustering in Section 4.3 reflects experiential interchangeability specifically, as distinct from a correlated confound, such as production era or energy level, that the residual-subtraction method does not disambiguate. These questions define the scope of the evaluation described in Section 6.5.
\section*{Acknowledgment}
The authors thank Last.fm for maintaining the public API that made this work possible and for their guidance on the research use of their data.

\section*{Data Availability and Ethical Use}
The listening data analyzed in this work was collected from the Last.fm public API and is used solely for this non-commercial research project. No raw or processed listening data is published or shared. Only aggregate statistics and derived findings are reported here. The data was handled on an anonymized basis, is retained as a single copy for the sole purpose of this research, and will be deleted on completion of the project. The authors contacted Last.fm regarding this collection and adhere to Last.fm's standard terms for academic and research use of its data. A copy of this work will be provided to Last.fm.

\bibliographystyle{IEEEtran}
\bibliography{references}
\end{document}